\documentclass[12pt]{iopart}
\usepackage{epsfig}
\usepackage{citesort}
\begin{document}
\newcommand{\no}{\nonumber\\}
\def\tgb{\mbox{$\tan{\beta}~$}}
\def\bsg{$b\to s \gamma$~}
\def\eps{$\varepsilon$~}
\def\epspeps{$\varepsilon^{\prime}/\varepsilon$~}
\def\Lsoft{${\cal L}_{SB}$~}
\def\mch{$m_{\chi^{\pm}}$~}
\def\mneu{$m_{\chi^{0}}$~}
\def\mglu{$m_{\tilde{g}}$~}
\def\stop{$m_{\tilde{t}}$~}
\def\mgrav{$m_{3/2}$~}
\def\Ibanez{Iba\~{n}ez~}
\def\Munoz{Mu\~{n}oz~}
\newcommand{\BXcenu}{B\rightarrow X_c e \nu}
\newcommand{\mub}{\mu_b}
\newcommand{\mb}{m_b}
\newcommand{\alphas}{\alpha_s}
\newcommand{\alphae}{\alpha_e}
\newcommand{\BRg}{{\rm BR}(B\to X_s \gamma)}
\newcommand{\BR}{{\rm BR}}
\newcommand{\Bsg}{B\to X_s \gamma}

\def\be{\begin{equation}}
\def\ee{\end{equation}}
\def\bea{\begin{eqnarray}}  
\def\eea{\end{eqnarray}}   
\def\etal{{\it et al.}}   
\def\eg{{\it e.g.}}
\def\ie{{\it i.e.}}
\def\Frac#1#2{\frac{\displaystyle{#1}}{\displaystyle{#2}}}
\def\lsim{\raise0.3ex\hbox{$\;<$\kern-0.75em\raise-1.1ex\hbox{$\sim\;$}}}
\def\gsim{\raise0.3ex\hbox{$\;>$\kern-0.75em\raise-1.1ex\hbox{$\sim\;$}}}
\renewcommand{\O}{{\cal O}}
\def\ap#1#2#3{     {\it Ann. Phys. (NY) }{\bf #1} (#2) #3}
\def\arnps#1#2#3{  {\it Ann. Rev. Nucl. Part. Sci. }{\bf #1} (#2) #3}
\def\npb#1#2#3{    {\it Nucl. Phys. }{\bf B #1} (#2) #3}
\def\npbps#1#2#3{    {\it Nucl. Phys. }(Proc. Suppl.){\bf B #1} (#2) #3}
\def\plb#1#2#3{    {\it Phys. Lett. }{\bf B #1} (#2) #3}
\def\prd#1#2#3{    {\it Phys. Rev. }{\bf D #1} (#2) #3}
\def\prep#1#2#3{   {\it Phys. Rep. }{\bf #1} (#2) #3}
\def\prl#1#2#3{    {\it Phys. Rev. Lett. }{\bf #1} (#2) #3}
\def\ptp#1#2#3{    {\it Prog. Theor. Phys. }{\bf #1} (#2) #3}
\def\rmp#1#2#3{    {\it Rev. Mod. Phys. }{\bf #1} (#2) #3}
\def\zpc#1#2#3{    {\it Zeit. f\"ur Physik }{\bf C #1} (#2) #3}
\def\mpla#1#2#3{   {\it Mod. Phys. Lett. }{\bf A #1} (#2) #3}
\def\sjnp#1#2#3{   {\it Sov. J. Nucl. Phys. }{\bf #1} (#2) #3}
\def\yf#1#2#3{     {\it Yad. Fiz. }{\bf #1} (#2) #3}
\def\nc#1#2#3{     {\it Nuovo Cim. }{\bf #1} (#2) #3}
\def\jetpl#1#2#3{  {\it JETP Lett. }{\bf #1} (#2) #3}
\def\ibid#1#2#3{   {\it ibid. }{\bf #1} (#2) #3}

\title[]{Supersymmetry predictions for $\varepsilon'/\varepsilon$}

\author{Shaaban Khalil}

\address{Centre for Theoretical Physics, University of Sussex, Brighton 
BN1 9QJ,~~~U.~K.}  

\address{Ain Shams University, Faculty of Science, Cairo, 11566, Egypt.}

\begin{abstract}
We study the predictions for the direct CP violation parameter \epspeps  in a class
of string--inspired model with non--universal soft $A$--terms. We show that the 
non--universality of the $A$--terms plays an important rule in enhancing the supersymmetric
contributions to the CP violation \epspeps. In particular, the supersymmetric contribution 
to \epspeps can be of order the KTeV result, $\varepsilon'/\varepsilon \simeq 10^{-3}$.

\end{abstract}


\section{Introduction}

The most recent results of \epspeps, which measures the size of the direct 
CP violation in $K_L \to \pi \pi$, reported by KTeV
\cite{CP1} and NA48 \cite{CP2}
lead to a world average
of Re~\epspeps =$(21.4\pm 4.0)\times 10^{-4}$~\cite{CP3}.
This result is higher than the Standard Model (SM) predictions \cite{epsp1},
opening the way to the interpretation that it may be a signal of
new physics beyond the SM. The SM predictions for \epspeps suffer from  large
theoretical uncertainties \cite{epsp2} such that one can not draw a definite conclusion
if this observed high value of \epspeps can be accommodated in the SM. In any case, one
may wonder if the supersymmetry (SUSY) can be responsible for enhancing \epspeps. 

In the minimal supersymmetric extension of the SM (MSSM) there is no way of generating
a siable SUSY contribution to \epspeps even if one assume that the SUSY CP violating
phases are large and the electric dipole moments (EDM) of the electron and neutron are
less than the experimental bounds due to the cancellation between the different
contributions. This is mainly due to the assumption of universal boundary conditions
of the soft-breaking terms \cite{GG,abel,khalil1,demir}. It has been shown that, 
without new flavor structure beyond the usual Yukawa couplings, general SUSY models with 
phases of the soft terms of order $\O(1)$ (but with a vanishing CKM phase 
$\delta_{\mathrm{CKM}}=0$)
can not give a sizeable contribution to the CP violating processes
\cite{abel,khalil1,demir,barr}. This means that the presence of non--universal soft breaking
terms besides large SUSY phases is crucial to enhance these CP violation effects.
In agreement with this, it has been explicitly shown
that contributions to $\varepsilon_K$ are small within the
dilaton--dominated SUSY breaking of the weakly coupled heterotic string
model~\cite{barr}, where $A$--terms as well as gaugino masses are universal. 
On the other hand, it is well--known that the strict universality in the soft
breaking sector is a strong assumption not realized in many supergravity and
string inspired models~\cite{ibanez1,ibanez2}.
All these arguments indicate not only that the presence of non--universal
soft terms can solve the problem of too large contributions to EDMs but also
that it allows for large SUSY contributions in CP violation experiments.
Hence, in this work we will follow this avenue and analyze the effects of
non--universal soft terms in CP violation in the $K$--system~\cite{khalil2,vives,emidio}.

\section{CP violation in minimal supergravity model}

It is well known that in SUSY models there are new possibilities for CP violation. 
In particular, the soft SUSY breaking terms contain several parameters that  may be 
complex, as can also be the $\mu$-parameter. 
In the minimal supergravity model there are
only two new CP-violating phases. This can be seen as  
follows. The parameters $M, A$ and $B$ and $\mu$ can be complex.
But of these four phases only two are physical. First, by an
R-rotation with R-charge $Q_R=1$ for lepton and quark superfields
and $Q_R=0$ for the vector and the Higgs superfields, the gaugino
mass parameter $M$ can be made real. Second, $B \mu$ can be made
real by a change of phase of the Higgs superfield. This ensures   
that the Higgs vacuum expectation values are real. The remaining
phases cannot be defined away and violate CP. One is in $A=A_0  
e^{i \phi_A}$ and the other in $B=B_0 e^{i\phi_B}$. The $\mu$
parameter then has a fixed phase $\mu=\mu_0 e^{-i\phi_B}$.
In any phase convention 
\be 
\phi_A= \mathrm{arg}(AM^*),  \hspace{3cm}
\phi_B= \mathrm{arg}(BM^*).
\ee
These phases can cause at one
loop level an electric dipole moment (EDM) for the quarks and
leptons, and therefore also for the neutron. It has been known for
a long time that in the SUSY models the contributions to the
neutron electric dipole moment are larger than the experimental
limit $6.3\times 10^{-26}$ e cm unless either the new `SUSY
phases' are tuned to be of order $10^{-3}$, or the SUSY masses are
of order a TeV. Such small phases can not generate sizable CP violation.
Also they constitute a fine tuning. This is known as `` SUSY CP problem".
However, in the last few it has been suggested that a natural   
cancellation mechanism exists whereby the electric dipole moment
of the neutron may be made small without such fine-tuning. In this case 
large SUSY phases are expected and still satisfy experimental bounds
on the values of EDM of the electron and neutron.

In this section  we will study the effect of these phases in CP violation 
observables as \eps and \epspeps. We assume that $\delta_{CKM}=0$ to maximize 
this effect. The value   
of the indirect CP violation in the Kaon decays, $\varepsilon$, is 
defined as $ \varepsilon = e^{i\frac{\pi}{4}}
{\rm Im} M_{12}/\sqrt{2} \Delta m_K,$
where $\Delta m_K= 2 {\rm Re} \langle K^0 \vert H_{eff} \vert \bar{K}^0
\rangle = 3.52 \times 10^{-15}$ GeV.
The amplitude
$M_{12}=\langle K^0 \vert H_{eff} \vert \bar{K}^0 \rangle$.
The relevant supersymmetric contributions to $K^0-\bar{K}^0$ are 
the gluino and the chargino contributions, (\ie, the transition
proceeds through box diagrams exchanging gluino-squarks and
chargino-squarks). It is usually expected that the gluino is the   
dominant contribution. However, as we will show, it is impossible
in the case of degenerate $A$-terms that the gluino gives any significant
contribution to $\varepsilon$ when the CKM matrix is taken to be real even with
large phase of $A$.
The amplitude of the gluino contribution is given in terms of the mass insertion 
$\delta_{AB}$ defined by $\delta_{AB} = \Delta_{AB}/\tilde{m}^2$ where
$\tilde{m}$ is an average sfermion mass and $\Delta$ is off-diagonal
terms in the sfermion mass matrices. The mass insertion to accomplish
the transition from $\tilde{d}_{iL}$ to $\tilde{d}_{jL}$ ($i,j$ are
flavor indices) is given by
\begin{eqnarray}
(\Delta^d_{LL})_{ij}&\simeq&-\frac{1}{8\pi^2}\left[\frac{K^{\dag}
(M_u^{diag})^2 K}{v^2 \sin^2\beta} \ln(\frac{M_{GUT}}{M_W})
\right](3\tilde{m}^2+\vert X \vert^2),
\\
(\Delta^d_{LR})_{ij}&\simeq&-\frac{1}{8\pi^2}\left[\frac{K^{\dag}
(M_u^{diag})^2 K\ M_d}{v^2 \sin^2\beta \cos\beta}   
\right] \ln(\frac{M_{GUT}}{M_W}) X,
\\
(\Delta^d_{RL})_{ij}&\simeq&-\frac{1}{8\pi^2}\left[\frac{M_d\ K^{\dag}
(M_u^{diag})^2 K }{v^2 \sin^2\beta \cos\beta}
\right] \ln(\frac{M_{GUT}}{M_W}) X,
\\
(\Delta^d_{RR})_{ij}&=&0,
\end{eqnarray}
where $X= A_d -\mu\ \tan\beta $. It is clear that $\Delta_{ij}$ in
general are complex due to the complexity of the CKM matrix, the
trilinear coupling $A$ and $\mu$ parameter. Here we assume the
vanishing of $\delta_{CKM}$ to analyze the effect of the SUSY
phases. We notice that $(\Delta^d_{LL})_{12}$ is proportional to
$\vert X \vert^2$ \ie, it is real and does not contribute to
$\varepsilon$ whatever the phase of $A$ is. Moreover, the values
of the $(\Delta^d_{LR})_{12}$ and $(\Delta^d_{RL})_{12}$ are
proportional to $m_s$ and $m_d$, hence they are quite small.
Indeed in this case we find the gluino contribution to
$\varepsilon$ is of order $10^{-6}$. 

For the chargino contribution the amplitude is given
by~\cite{branco}
\begin{eqnarray}
\hspace{-1cm}\langle K^0 \vert H_{eff} \vert \bar{K}^0 \rangle &=& -\Frac{G_F^2
M_W^2}{(2\pi)^2} (V_{td}^* V_{ts})^2 f_K^2 M_k \biggl[ \frac{1}{3} 
C_1(\mu) B_1(\mu)\no
&+& \bigl(\frac{M_k}{m_s(\mu) +m_d(\mu)}\bigr)^2 
\bigl(-\frac{5}{24} C_2(\mu) B_2(\mu) + \frac{1}{24} C_3(\mu)
B_3(\mu)\bigr)\biggr].
\end{eqnarray}
The complete expression for these function can be found in
Ref.~\cite{branco}. For low and modurate values of $\tan \beta$ the value
of $C_3$ is much smaller than $C_1$ since it is suppressed by the
ratio of $m_s$ to $M_W$. However, by neglecting the flavor mixing
in the squark mass matrix $C_1$ turned out to be exactly real~\cite{demir}.
The imaginary part of $C_1$ is associated to the size of the
intergenerational sfermion mixings, thus it is maximal for large
$\tan \beta$. In low $\tan \beta$ case, that we consider, the imaginary part
of $C_1$ is very small, and the gluino contribution is still the dominant
contribution $\varepsilon$. In particular, we found that \eps is of order 
$10^{-6}$, which is less than the experimental value $2.26\times 10^{-3}$.

Now we consider the effect of these two phases ($\phi_A$ and $\phi_{\mu}$)
on the direct CP violation parameter \epspeps. Similar to the case of indirect CP 
violation parameter \eps, in the gluino contribution the $L$--$L$ transitions are almost real
and the $L$--$R$ transitions are suppressed by two up Yukawa couplings and a down quark 
mass. Moreover, the analysis of chargino contribution is also the same as in the indirect CP 
violation. Even the experimental bounds on the branching ratio of $b \to s \gamma$ decay 
impose sever constraint on the LR transition. Therefore we do not find any significant SUSY
CP violation effect in \epspeps too.

\section{Non--universal soft terms and SUSY CP violation}

In the previous section, we have shown that CP violation effects are always very small 
in SUSY models with universal soft SUSY breaking terms. Recently, it has been shown that
the non--universality of $A$--terms is very
effective to generate large CP violation effects
\cite{abel,khalil1,khalil2,vives,masieromur,non-u}. In fact, the presence of
non--degenerate $A$--terms is essential for enhancing the gluino contributions
to $\varepsilon'/\varepsilon$ through large imaginary parts of the $L$--$R$
mass insertions, $\mathrm{Im}(\delta_{LR})_{12}$ and
$\mathrm{Im}(\delta_{RL})_{12}$.
These SUSY contributions can, indeed, account for a sizeable part of the
recently measured experimental value of $\varepsilon'/\varepsilon$
\cite{CP1,CP2}. In the following, we will present an explicit realization
of such mechanism in the framework of a type I superstring inspired SUSY
model. Within this model, it is possible to obtain non--universal soft
breaking terms, i.e. scalar masses, gaugino masses and trilinear
couplings. 

Type I string models contain nine--branes and three types of
five--branes ($5_a$, $a=1,2,3$).    
If we assume that the gauge group $SU(3) \times U(1)$ is on one of the branes (9--brane)
and the gauge group $SU(2)$ is on another brane ($5_1$--brane).
Chiral matter fields correspond to open strings spanning between
branes. Thus, they have non--vanishing
quantum numbers only for the gauge groups corresponding to
the branes between which the open string spans. For example, the chiral field corresponding 
to the open string between the $SU(3)$ and $SU(2)$ branes can have
non--trivial representations under both $SU(3)$ and $SU(2)$,
while the chiral field corresponding to the open string,
which starts and ends on the $SU(3)$--brane, should be
an $SU(2)$--singlet.
There is only one type of the open string which spans between the 9 and $5_1$--branes,
which we denote it as $C^{95_1}$.  
However, there are three types of open strings which could start and end on the
9--brane, that is, the $C^9_i$ sectors (i=1,2,3), which corresponding to the $i$-th 
complex compact dimension among the three complex dimensions. If we 
assign the three families to the different $C^9_i$ sectors we obtain non--universality
in the right--handed sector. In particular, we assign the $C^{9}_1$ sector
to the third family and $C^{9}_3$ and $C^{9}_2$ to the first and second families 
respectively. Under these assumption the soft SUSY breaking terms are obtained,
following the formulae in Ref.~\cite{ibanez}.
The gaugino masses are obtained
\begin{eqnarray}
\label{gaugino}
M_3 & = & M_1 = \sqrt 3 m_{3/2} \sin \theta\  e^{-i\alpha_S}, \\
M_2 & = &  \sqrt 3 m_{3/2} \cos \theta\ \Theta_1 e^{-i\alpha_1}. 
\end{eqnarray}
While the $A$-terms are obtained as
\begin{equation}
A_{C_1^9}= -\sqrt 3 m_{3/2} \sin \theta\ e^{-i\alpha_S}=-M_3,
\label{A-C1}
\end{equation}
for the coupling including $C_1^{9}$, i.e. the third family,
\begin{equation}
A_{C_2^9}= -\sqrt 3 m_{3/2}(\sin \theta\ e^{-i\alpha_S}+
\cos \theta\ (\Theta_1 e^{-i\alpha_1}- \Theta_2 e^{-i\alpha_2})),
\label{A-C2}
\end{equation}
for the coupling including $C_2^{9}$, i.e. the second
family  and
\begin{equation}
\label{A-C3}   
A_{C_3^9}= -\sqrt 3 m_{3/2}(\sin \theta\ e^{-i\alpha_S}+
\cos \theta\ (\Theta_1 e^{-i\alpha_1}- \Theta_3 e^{-i\alpha_3})),
\end{equation}
for the coupling including $C_3^{9}$, i.e. the first family.
Here $m_{3/2}$ is the gravitino mass, $\alpha_S$ and $\alpha_i$ are
the CP phases of the F-terms of the dilaton field $S$ and
the three moduli fields $T_i$, and $\theta$ and $\Theta_i$ are
goldstino angles, and we have the constraint, $\sum \Theta_i^2=1$.
Thus, if quark fields correspond to different
$C_i^9$ sectors, we have non--universal A--terms.
Then we obtain the following A--matrix for both of the
up and down sectors,
\begin{eqnarray}
A= \left(  
\begin{array}{ccc}
A_{C^9_3}  & A_{C^9_2} & A_{C^9_1} \\ A_{C^9_3} & A_{C^9_2} &
A_{C^9_1} \\ A_{C^9_3} & A_{C^9_2} & A_{C^9_1}
\end{array}
\right) \label{A-1}.
\end{eqnarray}
Note that the non--universality appears
only for the right--handed sector.
The trilinear SUSY breaking matrix, $(Y^A)_{ij}=(Y)_{ij}(A)_{ij}$,
itself is obtained
\begin{equation}
\label{trilinear}
Y^A = \left(\begin{array}{ccc}
 &  &  \\  & Y_{ij} &  \\  &  & \end{array}
\right) \cdot
\left(\begin{array}{ccc}
A_{C^9_3} & 0 & 0 \\ 0 & A_{C^9_2} & 0 \\ 0 & 0 & A_{C^9_1} \end{array}
\right),   
\end{equation}
in matrix notation. In addition, soft scalar masses for quark doublets and
the Higgs fields are obtained,
\begin{equation}
\label{doublets}
m_{C^{95_1}}^2=m_{3/2}^2(1-\Frac{3}{2} \cos^2 \theta\ (1-
\Theta_1^2)).
\end{equation}  
The soft scalar masses for quark singlets are obtained as
\begin{equation}
\label{singlets}
m_{C_i^9}^2=m_{3/2}^2(1-3\cos^2 \theta\ \Theta^2_i),
\end{equation}
if it corresponds to  the $C_i^{9}$ sector.

In models with non-degenerate $A$--terms we have to fix the Yukawa
matrices to completely specify the model. In fact, with universal
$A$--terms the textures of the Yukawa matrices at GUT scale affect
the physics at EW scale only through the quark masses and usual
CKM matrix, since the extra parameters contained in the Yukawa
matrices can be eliminated by unitary fields transformations. This
is no longer true with non-degenerate $A$--terms. 
Here, we choose our Yukawa texture to be
\be
\hspace{-1.5cm}Y^u=\frac{1}{v\cos{\beta}} {\rm diag}\left(
m_u,m_c,m_t\right)~,~~
Y^d=\frac{1}{v\sin{\beta}}  K^{\dagger} \cdot {\rm diag }
\left(m_d, m_s, m_b\right) \cdot  K
\ee
where $K$ is the CKM matrix. In this case one find that the mass inserion $\delta^d_{LR}$
can be written as~\cite{vives}
\begin{eqnarray}
\label{DLR}
\hspace{-1.5cm}(\delta_{LR}^{(d)})_{i j}&=& \frac{1}{m^2_{\tilde{q}}}\ m_i\ \Big(
\delta_{ij}\ (c_{A} A_{C^9_3}^*\ +\ c_{\tilde{g}}\ m_{\tilde{g}}^* -\ 
\mu e^{i\varphi_{\mu}} \tan\beta ) \nonumber \\
&+&K_{i 2}\ K^*_{j 2}\ c_{A}\ ( A_{C^9_2}^* - A_{C^9_3}^* ) +
K_{i 3}\ K^*_{j 3}\ c_{A}\ ( A_{C^9_1}^* - A_{C^9_3}^* ) \Big)
\end{eqnarray}
where $m^2_{\tilde{q}}$ is an average squark mass and $m_i$ the quark mass.
This expression shows the main effects of the non--universal $A$--terms.
In the first place, we can see that the diagonal elements are still very
similar to the universal $A$--terms situation. Apart of the usual scaling with the
quark mass, these flavor--diagonal mass insertions receive
dominant contributions from the corresponding $A_{C^9_i}$ terms
plus an approximately equal contribution from gluino to all three generations
and an identical $\mu$ term contribution. Hence, given that the gluino
RG effects are dominant, also the phases of 
these terms tend to align with the gluino phase, as in the minimal supergravity. 
Therefore, EDM bounds constrain mainly the relative phase between $\mu$ and gluino
(or chargino) and give a relatively weaker constraint to the relative
phase between $A_{C^9_3}$ (the first generation $A$--term) and the relevant
gaugino. Effects of different $A_{C^9_i}$ in these elements are suppressed by squared CKM
mixing angles. However, flavor--off--diagonal elements are completely new
in this model. They do not receive significant contributions from gluino
nor from $\mu$ and so their phases are still determined by the $A_{C^9_i}$
phases and, in principle, they do not directly contribute to EDMs .

\begin{figure}
\begin{center}
\epsfxsize = 11cm
\epsffile{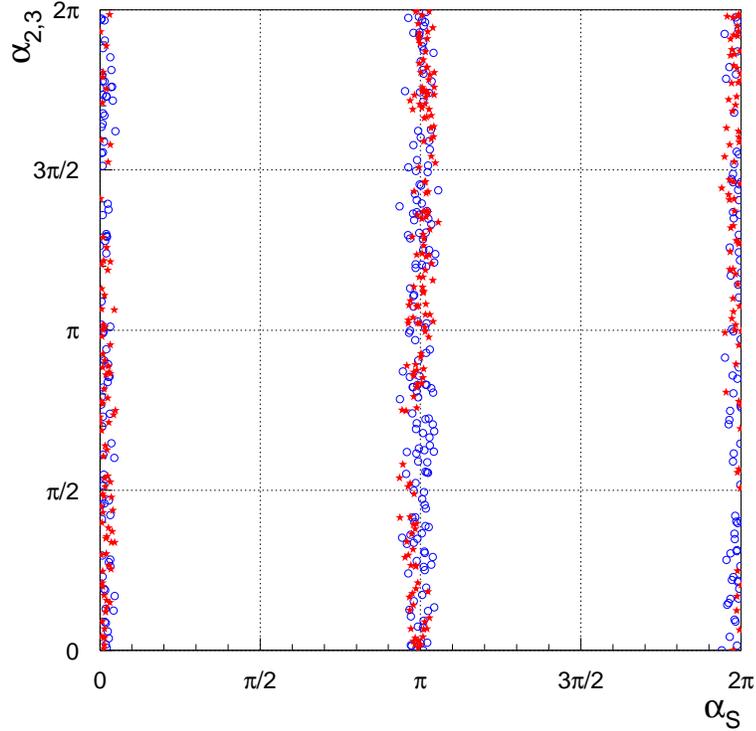}
\leavevmode
\end{center}
\caption{Allowed values for $\alpha_2$--$\alpha_S$ (open blue circles) and
$\alpha_3$--$\alpha_S$ (red stars)}
\label{scat}
\end{figure}

In figure \ref{scat} we show the allowed values for $\alpha_S$, $\alpha_2$
and $\alpha_3$ assuming $\alpha_1=\varphi_\mu=0$. We 
have imposed the EDM, \eps and $b \to s \gamma$ bounds with the usual bounds on SUSY 
masses. We can see that, similarly to the minimal supergravity, $\varphi_\mu$ is
constrained to be very close to the gluino and chargino phases
(in the plot $\alpha_S \simeq 0, \pi$), but $\alpha_2$ and
$\alpha_3$ are completely unconstrained.

\begin{figure}
\begin{center}
\epsfxsize = 11cm
\epsffile{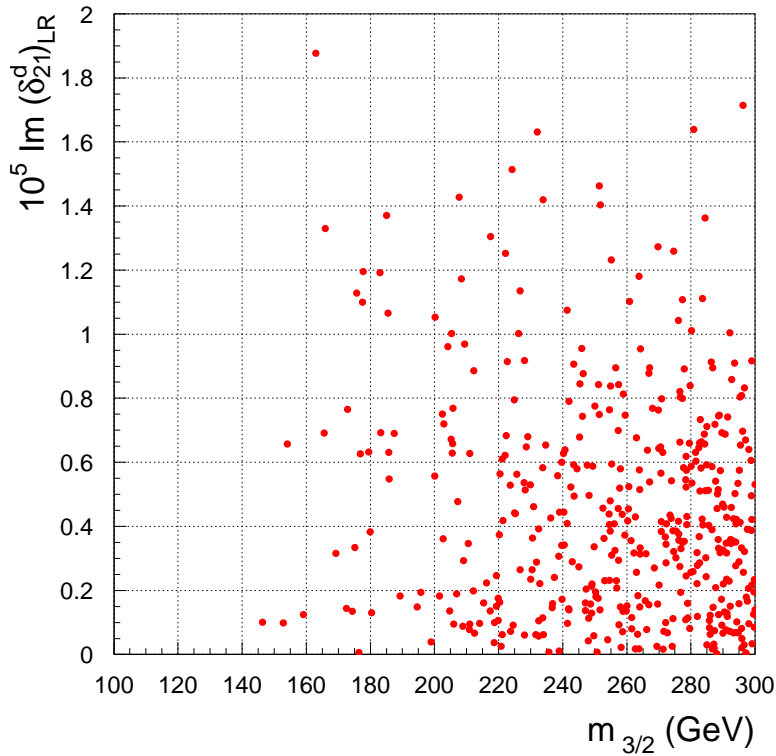}
\leavevmode
\end{center}
\caption{$(\delta_{LR}^{(d)})_{2 1}$ versus $m_{\tilde{g}}$ for experimentally
allowed regions of the SUSY parameter space}
\label{eps'}
\end{figure}

Finally, in figure \ref{eps'} we show the values of
$Im (\delta^{(d)}_{LR})_{2 1}$ versus the gluino mass in the same regions of
parameter space and with the same constraints as in figure \ref{scat}. 
As we can see due to the effect of the off-diagonal phases a large percentage of points 
are above or close to $1 \times 10^{-5}$, hence, sizeable supersymmetric contribution to
$\varepsilon^\prime/\varepsilon$ can be expected in the presence of
non-universal $A$--terms. 

\section{Conclusions}

Non--universal Supersymmetry soft breaking terms are a natural consequence
in many supergravity or string inspired SUSY models. Moreover, the
non--universality of the $A$--terms has a significant effect in the CP violation.
We have shown that in these models a sizeable supersymmetric contribution to CP 
observables \epspeps and  \eps can be easily obtained. 

\section*{ Acknowledgement}
I would like to thank S. Barr, T. Kobayashi, A. Masiero and O. Vives for their
collaboration in this project. I also would like to thank the organizers for such
a nice and stimulating atmosphere in which the workshop took place.\\




\begin{thebibliography}{99}

\bibitem{CP1}
A. Alavi--Harati {\it et al.} (KTeV Coll.), \prl{83}{1999}{22}.
\bibitem{CP2} 
V. Fanti {\it et al.} (NA48 Coll.), \plb{465}{1999}{335}.
\bibitem{CP3}
G. D'Agostini, hep-ex/9910036.
\bibitem{epsp1}
A. Buras, M. Jamin, and M.E: Lautenbacher, \npb{408}{1993}{209};
M. Ciuchini, E. Franco, G. Martinelli, and L. Reina, \npb{415}{1994}{403};
S. Bosh, A.J. Buras, M. Gorbahn, S. Jager, M. Jamin,
M.E. Lautenbacher, and L. Silvestrini, \npb{565}{2000}{3};
M. Ciuchini, E. Franco, L. Giusti, V. Lubicz, and
G. Martinelli, hep-ph/9910237; M. Jamin, hep-ph/9911390.
\bibitem{epsp2}
S. Bertolini, M. Fabbrichesi, and J.O. Eeg,  {\it Rev. Mod. Phys.}{\bf 72} (2000) 65;
T. Hambye, G.O. Kohler, E.A. Paschos, and P.H. Soldan,\npb{564}{2000}{391};
J.Bijnens, and J.Prades, JHEP 01, (1999) 023;
E. Pallante and A. Pich, \prl{84}{2000}{2568}.
\bibitem{GG}
E. Gabrielli and G.F. Giudice, \npb{433}{1995}{3};
Erratum-ibid. {\bf B~507} (1997) 549.
\bibitem{abel}
S.~Abel and J.~Frere, \prd{55}{1997}{1623}.
\bibitem{khalil1}
S.~Khalil, T.~Kobayashi, and A.~Masiero, \prd{60}{1999}{075003}.
\bibitem{demir}
D.A.~Demir, A.~Masiero, and O.~Vives, \plb{479}{2000}{230};
D.~A.~Demir, A.~Masiero, and O.~Vives, \prd{61}{2000}{075009}.
\bibitem{barr}
S.~Barr and S.~Khalil, \prd{61}{2000}{035005}.
\bibitem{ibanez1}
A. Brignole, L. E. \Ibanez, and C. \Munoz, \npb{422}{1994}{125},
Erratum-ibid. {\bf B 436}~(1995) 747.
\bibitem{ibanez2}
L. E. \Ibanez, C. \Munoz, and S. Rigolin, \npb{553}{1999}{43}.
\bibitem{khalil2}
S.~Khalil and T.~Kobayashi, \plb{460}{1999}{341}.
\bibitem{vives}
S. Khalil, T. Kobayashi, and O. Vives,  \npb{580}{2000}{275}. 
\bibitem{emidio}
E. Gabrielli, S. Khalil, E. Torrente-Lujan, hep-ph/0005303. 
\bibitem{branco}
G.C.~Branco, G.C.~Cho, Y.~Kizukuri and N.~Oshimo,
\npb{449}{95}{483}.
\bibitem{masieromur}
A. Masiero and H. Murayama, \prl{83}{1999}{907}.
\bibitem{non-u}
R.~Barbieri, R.~Contino and A.~Strumia, \npb{578}{2000}{153};\\
K.~Babu, B.~Dutta and R.N.~Mohapatra, \prd{61}{2000}{091701}.

\end{thebibliography}
\end{document}